\documentclass[reprint,3p]{elsarticle}

\usepackage{subcaption}
\usepackage{color}

\usepackage{amsmath,amssymb}

\usepackage{lineno,hyperref}

\begin{document}

\begin{frontmatter}

\title{Modelling Microbial Fuel Cells using lattice
Boltzmann methods}


%
%

\author[label1]{Michail-Antisthenis Tsompanas\corref{cor1}}
\address[label1]{Unconventional Computing Centre, University of the West of England, Bristol BS16 1QY, UK\fnref{label1}}
\cortext[cor1]{Corresponding author}
\ead{antisthenis.tsompanas@uwe.ac.uk}

\author[label1]{Andrew Adamatzky}

\author[label2]{Ioannis Ieropoulos}
\address[label2]{Bristol BioEnergy Centre, University of the West of England, Bristol BS16 1QY, UK\fnref{label2}}

\author[label1]{Neil Phillips}

\author[label4]{Georgios Ch. Sirakoulis}
\address[label4]{Department of Electrical and Computer Engineering, Democritus University of Thrace, Xanthi 67100, Greece\fnref{label4}}

\author[label2]{John Greenman}


\begin{abstract}
An accurate modelling of bio-electrochemical processes that govern Microbial Fuel Cells (MFCs) and mapping their behaviour according to several parameters will enhance the development of MFC technology and enable their successful implementation in well defined applications. The geometry of the electrodes is among key parameters determining efficiency of MFCs due to the formation of a biofilm of anodophilic bacteria on the anode electrode, which is a decisive factor for the functionality of the device. We simulate the bio-electrochemical processes in an MFC while taking into account the geometry of the electrodes. Namely, lattice Boltzmann methods are used to simulate the fluid dynamics and the advection--diffusion phenomena in the anode compartment. The model is verified on voltage and current outputs of a single MFC derived from laboratory experiments under continuous flow.
\end{abstract}

\begin{keyword}
microbial fuel cells,  lattice Boltzmann, modelling, agent-based model, bio-electrochemical processes
\end{keyword}

\end{frontmatter}

\section*{Introduction}
Due to improvements in power and current outputs achieved in the last decades \cite{ieropoulos2010improved}, studies in Microbial Fuel Cells (MFCs) have intensified \cite{santoro2017microbial}. Despite the increased interest in this field, the majority of research work focuses on practical experimental analysis. On the other hand, just a handful of studies \cite{ortiz2015developments} have steered towards mathematically describing and simulating the bio-electrochemical processes that generate the electrical output of these devices. Arguably, the intrinsic complexity of these systems requires highly elaborative procedures to produce accurate models. However, the advantages (in time and cost saving terms) acquired by an accurate MFC model has motivated researchers \cite{korth2015framework,recio2016combined, tsompanas2017cellular} to exert more effort towards this direction.

The utilisation of porous electrodes in continuous flow MFCs, which is realised as best practice due to higher outputs \cite{kim2012porous}, and the observation that the macrostructure of the anode, its material and its inhomogeneities have a significant impact \cite{Merkey2012,gajda2015simultaneous}, prompted the employment of the lattice Boltzmann method (LBM) in the framework of the MFC model presented here. LBM is a numerical tool based on mesoscopic phenomena and is able to model fluid dynamics and transport (advection--diffusion) phenomena in fluid systems \cite{Chopard2002}. In contrast to its precursor, lattice gas automata that studies the individual molecular dynamics of fluids or gases --in a microscopic scale--, LBM simulates transport phenomena by studying the evolution of ensembles of molecules --in a mesoscopic scale. Nonetheless, the basic principles of microscopic physics are retained due to the utilisation of the ensemble averaged distribution function as the primary variable \cite{Chopard2002}. LBMs have an inherent parallel nature that is advantageous when implementing the algorithm in specialised parallel hardware. Moreover, the lattices used can trivially illustrate complex geometries compared with other computation methods.

The ultimate goal of the proposed model is to simulate in an efficient way the behaviour of an MFC device, namely the voltage and current output. That is achieved by analysing in the macroscopic and mesoscopic scales several phenomena occurring in the anode half-cell of an MFC --assuming that the cathode performance is not limiting. Specifically, fluid dynamics, transport phenomena, bio-electrochemical processes and biofilm formation are all taken into consideration and, thus, simulated by separate procedures, that constitute the proposed model whilst interacting with each other. The formation of a biofilm of anodophilic bacteria on the anode electrode is a decisive factor for the functionality of the device. Fluid dynamics and transport phenomena were simulated using two different LBMs, one providing the velocity of the fluid throughout the porous electrode positioned in the anode compartment and the other calculating the concentration of chemicals --given the velocity field acquired from the first LBM.

The use of LBMs, in addition to providing the ability to represent inhomogeneities or complex geometries of electrodes, empowers the faster execution of critical calculations that can be considered as the bottleneck of previous studies that have studied these phenomena \cite{picioreanu2010model}. That is due to the numerous iterations needed for the system to reach an equilibrium in the procedures of fluid dynamics and transport phenomena and the frequency at which these two procedures have to be updated, namely every time the biofilm formation is updated.

To support the aforementioned statement, note that the deciding chemical, physical and biological processes occurring in an MFC need to be studied across a wide span of time scales \cite{von2009three,bottero2013biofilm}. Although the time for a system to reach an equilibrium concerning its fluid dynamics and transport of chemicals, is in the period of some seconds, the evolution --or even a small change of the formation-- of a biofilm is noted after some hours or days. As a result, when modelling the evolution in the behaviour of a system that includes such a complicated collection of interdisciplinary processes, the faster relaxing (reaching equilibrium) procedures, here fluid dynamics and advection-diffusion, will have to be revised every time the slower updating process provides slightly different outputs. That introduces a bottleneck in the efficient execution of such detailed models. 

\section{Previous work}
\label{sec2}

LBMs have been extensively studied and claimed to be a viable alternative to other computational methods \cite{Chopard2002}. This method can be used for different modelling purposes, such as reaction-diffusion systems, complex fluid dynamics or wave propagation \cite{Chopard2002,wolf2004lattice}. LBM is considering a mesoscale statistical representation of groups of molecules and is divided into two processes, the collision and the propagation of these groups. The collision operator characterising each LBM, leads to the relaxation of the distributions of the groups to a local equilibrium which is the result of a set of parameters like the velocity, the concentration (considered as conserved quantities) and external forces.

The robustness of different types and variations of LBMs have been studied for several problems. For instance, in \cite{pan2006evaluation}, the fluid dynamics through complex geometries, namely porous media, was evaluated for several types of LBM and with different boundary conditions. In \cite{prasianakis2013simulation}, a three-dimensional LBM was suggested, implementing twenty seven characteristic velocities to model the flow field in porous media. The authors paralleled the results from their model with the ones from other numerical models and measurements from experiments.

In addition to fluid dynamics, LBM has successfully tackled the solution of advection--diffusion equations with high accuracy \cite{chopard2009lattice}. Moreover, the ability of LBM to model nonlinear convection--diffusion phenomena with anisotropic diffusion have been presented in \cite{shi2011lattice}, where the modelling of several one- and two-dimensional paradigms was investigated in detail. In \cite{li2017lattice} a comparison of different lattices has been executed to analyse the robustness and accuracy of them. Also, the impact that the chosen scheme of boundary conditions has in the accuracy of the model was studied \cite{huang2015boundary}.  


Moreover, researchers have previously employed the LBM in the field of conventional fuel cells. For instance, in \cite{joshi2007lattice} the diffusion of gases --that are characterised by different molecular weights-- were modelled in complex porous media, specifically in a solid oxide fuel cell (SOFC) anode. The authors in \cite{wang2006modeling} studied the flow field in a three-dimensional scheme of a section of a serpentine channel and in a two-dimensional scheme of a channel filled or partially filled with a porous medium, taking particular interest in the boundary conditions and the simulated time periods. Moreover, in \cite{dang2016pore} a multi-component flow model was presented to analyse the transport of a mixture of gases in a SOFC anode through irregular porous media. The impact of the micro-structure of the electrode, the dimensionless reaction flux and fuel composition to the fuel cell outputs were studied. Also, the effect that the micro-structure of the anode has on the distribution and deposition of carbon and, thus, to the performance of the cell was studied \cite{xu2016lattice}.

In addition, LBMs have been used in three-dimensional simulation of biofilm growth and evolution with respect to the flow dynamics and the advection--diffusion of chemicals through porous media \cite{von2009three,pintelon2012effect}. Two types of biofilms were studied, with zero permeability \cite{von2009three} and non-zero permeability \cite{pintelon2012effect}. The main difference in the work presented here with the aforementioned, is that here the function of the anode compartment of an MFC is simulated, namely, the bio-electrochemical reactions were studied in addition to the flow dynamics, transport phenomena and biofilm formation. Nonetheless, the impact that the bio-electrochemical phenomena have towards all the other procedures were included to the model. Keeping in mind the ultimate goal of an enhanced performance of the model, the presented model is based on a unified discrete grid where all procedures of the model are calculated. On the other hand, previous works \cite{von2009three,pintelon2012effect} have used LBMs based on discrete grids, whereas the simulation of biofilm is tackled with an individual-based model \cite{picioreanu2004particle} utilising continuous Cartesian dimensions. This incoherency in the procedures of the model will pose difficulties while implementing the model efficiently in software (or indeed hardware).

In the present work the selection of using a two-dimensional grid, rather than a three-dimensional similar to previous studies, certainly produces a lower accuracy, as claimed in \cite{von2009three,pintelon2012effect}. However, the use of a two-dimensional grid enables a faster execution of the model, whilst scaling up the grid to three dimensions is trivial. Moreover, the utilisation of the same grid for all the procedures of the model, is an additional advantage when advancing towards three dimensions.
%
%
%
%
%
%
%
%


As mentioned before, a great advantage of LBMs is their capacity of significant acceleration which is straightforward, due to the inherently parallel characteristics of LBM. The parallel implementation of the algorithm of the model can be realised with advanced programming languages compatible with multi-core processors \cite{bacsaugaouglu2016computational} or specialised hardware --like Field Programmable Gate Arrays (FPGAs) \cite{4439254,903392} or Graphics Processing Units (GPUs) \cite{5362489}.

The biofilm formation is a process that it is not yet fully decoded by scientific research. The accurate modelling of biofilms is a quite demanding task, given that the behaviour of living, self-organising organisms has to be taken into consideration and the experimental justification of simulated results is challenging. However, several examples of previous work aimed to explain and model the underlying mechanics of the evolution of the biofilm, like attachment, growth, decay, lysis and detachment of biomass \cite{bottero2013biofilm,picioreanu2004particle,stewart2002pore,bol20093d}. A simple agent-based model 
that can simulate the basic procedures of attachment and development was used here. Note here, that despite the simplicity of the model, it is capable of robustly formulating the under study processes.

Further advantages of the LBM model are that a detailed analysis of mass transfer can be carried out for the actual micro-structure of the anode electrode and the cell's performance can be judged by calculating the concentration polarisation. Complex geometries can be easily handled and both continuum and non-continuum diffusion through the pores can be modelled. Because of the level of detail, the effect of anode geometry on voltage polarisation can be directly studied and the geometry can be optimised to provide enhanced performance. In fact, being able to model an anodic micro-structure may be advantageous in understanding 3D porous electrode behaviour of real MFCs and their biofilms.


\section{Methods}
\label{sec3}

To approximate the behaviour of a MFC the model simulates the formation of a biofilm on the electrode surface, the fluid dynamics and the advection--diffusion phenomena throughout the anode compartment and calculates outputs of the MFC based on local concentration of chemicals and electrochemical equations. Two LBMs were used. 
One LBM represents the dynamics of fluids inside the anode compartment and another LBM represents advection-diffusion equation of chemicals. The model runs as follows: 
\begin{enumerate}
    \item Formation of the electrode geometry and initialisation of the parameters.
    \item Calculation of fluid dynamics with LBM (first LBM model).
    \item Calculation of the advection--diffusion equation with LBM (second LBM model).
    \item Calculation of MFC outputs (bio-electrochemical model).
    \item Biofilm formation.
    \item Go to step (2).
\end{enumerate}

\subsection{Lattice Boltzmann Method}
\label{sec2}

\begin{figure}[!tbp]
    \centering
       \includegraphics[width=0.6\linewidth]{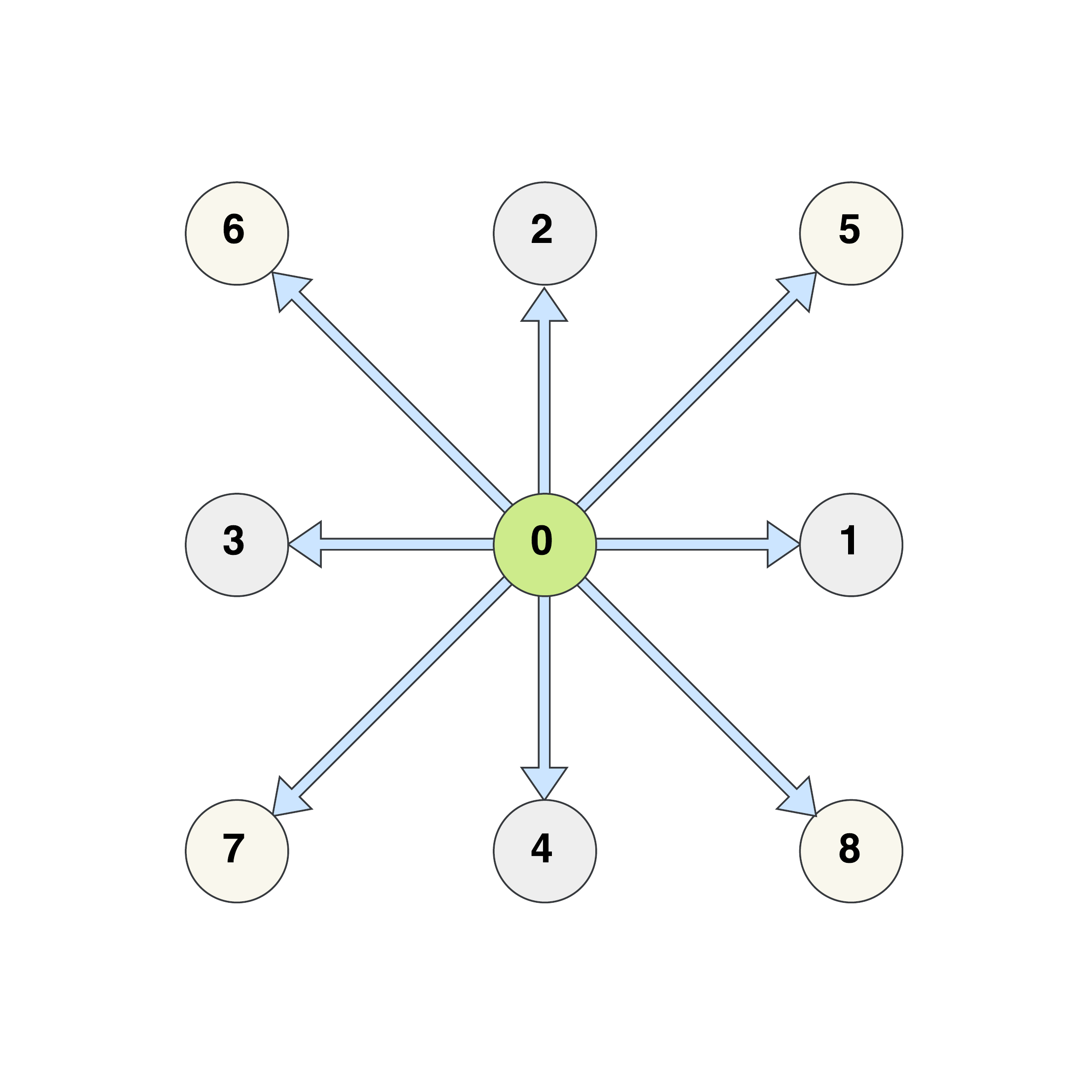}
    \caption{The D2Q9 lattice.}
    \label{D2Q9}
\end{figure}

A LBM is characterised by three main factors: the lattice, the equilibrium equation and the kinetic equation. We adopt the Bhatnagar-Gross-Krook (BGK) LBM framework~\cite{Qian92} with a D2Q9 lattice (Fig. \ref{D2Q9}), based on a two-dimensional lattice with nine characteristic velocities: 

\begin{equation}
c_{0}=(0,0)
\label{speed1}
\end{equation}

\begin{equation}
c_{1,3}=(\pm 1,0) \quad c_{2,4}=(0,\pm 1)
\label{speed2}
\end{equation}

\begin{equation}
c_{5,6,7,8}=(\pm 1,\pm 1)
\label{speed3}
\end{equation}

The particle velocities distribution is governed by the evolution equation which takes into account propagation and collision processes: 

\begin{equation}
F_i(x+c_i\Delta t, t +\Delta t) - F_i(x,t) = - \frac{\Delta t}{\tau} \bigg[ F_i(x,t)-F_i^{(0)}(x,t) \bigg]
\label{kinet}
\end{equation}

\noindent where $\tau$ is the dimensionless relaxation time, $t$ is a simulation time step, $x$ is a site in the lattice, and $i$ is an index of the characteristic velocity. 
%
%
%
$F_i^{(0)}(x,t)$ is the equilibrium distribution function:

\begin{equation}
F_i^{(0)}(\rho , j)= \frac{W_i}{\rho_0} \rho \Bigg[ 1 + 3\frac{c_i\cdot u}{c^2}+\frac{9}{2}\frac{(c_i \cdot u)^2}{c^4}-\frac{3}{2}\frac{u^2}{c^2} \Bigg]
\label{equi0}
\end{equation}

where $c$ is a model constant relating to the lattice speed ($c=\frac{dx}{dt*\sqrt{3}}$), $c_i$ are defined in Eqs. (\ref{speed1}-\ref{speed3}), $u$ is the velocity flow, $\rho$ is the mass density and $\frac{W_i}{\rho_0}$ is a weight factor equal to:  
\begin{equation}
\frac{W_i}{\rho_0}=\frac{4}{9}  \; \textnormal{,    for} \;\; i=0
\label{equi1}
\end{equation}

\begin{equation}
\frac{W_i}{\rho_0}=\frac{1}{9}  \; \textnormal{,    for} \;\; i=1,2,3,4
\label{equi2}
\end{equation}

\begin{equation}
\frac{W_i}{\rho_0}=\frac{1}{36}  \; \textnormal{,    for} \;\; i=5,6,7,8.
\label{equi3}
\end{equation}
%
%
%
%

\subsubsection{Fluid dynamics}

Fluid dynamics are calculated by LBM with boundaries defined by the compartment physical boundaries (i.e. plastic container), the anode electrode geometry and the dynamic biofilm formation. 
The method is based on the BGK model, see Eqs. (\ref{speed1} -- \ref{equi3}). 
This method is utilising bounce-back boundary conditions \cite{Chopard2002} and has a second order accuracy for fluid dynamics. 
The resulting mass density ($\rho$) and momentum density ($j$) determine the flow field as follows:

\begin{equation}
\rho(x,t)=\sum_{i}F_i(x,t)
\label{mass}
\end{equation}

\begin{equation}
j(x,t)=\rho(x,t)u(x,t)=\sum_{i}c_i F_i(x,t)
\label{momentum}
\end{equation}

\begin{figure}[!tbp]
    \centering
       \includegraphics[width=0.9\linewidth]{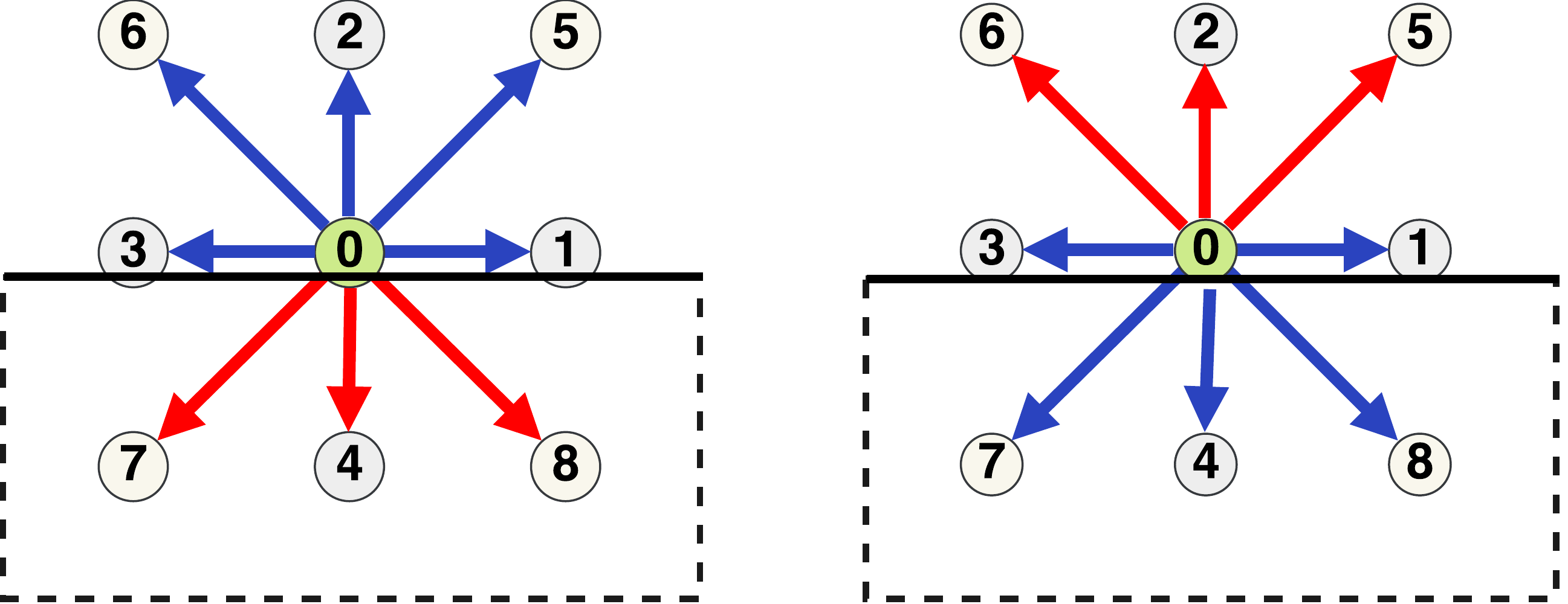}
    \caption{The bounce-back boundaries principle in a D2Q9 lattice.}
    \label{boundar}
\end{figure}

On the boundaries the bounce-back principle dictates that the evolution equation is not calculated as in Eq. (\ref{kinet}). However, the collision process is updated depending on the configuration of the boundaries and the available space. For instance, in the case of a boundary placed in the south direction of the lattice (Fig. \ref{boundar}) the following equations apply:

\begin{equation}
F_5(x,t)=F_7(x,t), \quad F_2(x,t)=F_4(x,t), \quad F_6(x,t)=F_8(x,t).
\label{momentum1}
\end{equation}

\subsubsection{Advection--diffusion equations}

The advection--diffusion equations are solved using another BGK LBM, where the flow field and geometry of electrodes are used as inputs.
The boundary conditions are considered as Dirichlet boundaries for the input and output of the domain (with a given concentration amount) and as Neumann for the solid boundaries (with a set zero-flux of concentration towards them) \cite{zhang2012}. 
The evolution equation for the distribution function is similar to Eq. (\ref{kinet}) 
%
%
%
%
(note the different notation of the distribution function used -- $G_i$ rather than $F_I$ -- to distinguish the two separate LBMs):

\begin{equation}
G_i(x+c_i\Delta t, t +\Delta t) - G_i(x,t) = - \frac{\Delta t}{\tau_D} \bigg[ G_i(x,t)-G_i^{(0)}(x,t) \bigg] 
\label{kinet_dif}
\end{equation}

\noindent where $G_i^{(0)}(x,t)$ is the equilibrium distribution function, which is calculated as in Eq. (\ref{equi0}) and $W_i$ are the weights defined by Eqs. (\ref{equi1})-(\ref{equi3}). 
Note that 
the velocity ($u$) is considered as a constant for the advection--diffusion simulation and provided by the first LBM procedure.
%
%
%
%

The resulting local concentration ($C_s$) which represents the concentration field is calculated in the following : 

\begin{equation}
C_s(x,t)=\sum_{i}G_i(x,t)
\label{mass1}
\end{equation}

\subsection{Bio-electrochemical model}


The mechanism for the transfer of electrons to the anode is assumed to incorporate the existence of an intracellular mediator and the interconnection of the biomass with the electrode by nanowires or by direct contact \cite{recio2016combined,pinto2010two}. In the simulated experiments the total concentration of the intracellular mediator is regarded as constant and analogous to the biomass concentration. After calculating the equilibrium of fluid dynamics and the concentration of the chemical species, the output of MFC is studied using the following equations: 

\begin{equation}
M_{total}=M_{red}+M_{ox}
\label{med1}
\end{equation}

\begin{equation}
M_{ox}(x,t+1)=M_{ox}(x,t)-Y q_a+\gamma \frac{I_{cell}}{m F} \frac{1}{V_a C^{bio}} 
\label{med2}
\end{equation}

\noindent where $M_{total}$ is the total amount of intracellular mediator, $M_{red}$ is the reduced form, while $M_{ox}$ is the oxidised form. $Y$ is the mediator yield granted by the utilisation of the substrate, $\gamma$ is the mediator molar mass, $m$ is the fraction of electrons provided during reduction of the intracellular mediator. $F$ is the Faraday constant, $V_a$ is the anode compartment volume, $C^{bio}$ is the concentration of biomass. $I_{cell}$ is the total current produced and, thus, flowing through the device and $q_a$ is the consumption rate of the substrate.

The double Monod equation is used to determine 
the consumption rate of the substrate:

\begin{equation}
q_a=q_{max} \frac{C_s}{C_s+K_s} \frac{M_{ox}}{M_{ox}+K_{M_{ox}}} 
\label{monod}
\end{equation}

\noindent where $C_s$ is the concentration of the substrate, $K_s$ is the half saturation constant for substrate, $K_{M_{ox}}$ is the half saturation constant for oxidised mediator and $q_{max}$ is the maximum consumption rate.

The MFC as an electrochemical system suffers from losses, defined elsewhere as polarisation or overpotential, caused by the rates of the reactions (activation overpotential), the ohmic resistances of the material used to build the MFC (ohmic overpotential) and the limitations posed on the mass transfer or kinetic phenomena (concentration overpotential). As a result, considering Kirchhoff's voltage law and Ohm's law the following equation correlates the overpotentials, the values of resistance and the produced current.

\begin{equation}
I_{cell} R_{ext} = E_0 - I_{cell} R_{int} - n_{conc} - n_{act}
\label{con_ov}
\end{equation}

\noindent where $R_{ext}$ and $R_{int}$ are the external (load) and the internal resistances, $E_0$ is the maximum produced voltage (or the open circuit voltage), $n_{conc}$ and $n_{act}$ are the concentration and activation overpotentials, respectively. The concentration overpotential based on the Nerst equation is defined by:

\begin{equation}
n_{conc}=\frac{RT}{F} ln \Bigg( \frac{M_{total}}{M_{red}} \Bigg) 
\label{con_ov}
\end{equation}


\noindent where $R$ is the universal gas constant and $T$ is the temperature; while, the activation overpotential can be calculated by an approximation of the Butler-Volmer equation:

\begin{equation}
n_{act}=\frac{I_{cell}}{A_a I_0} \frac{RT}{mF} \Bigg( \frac{M_{ox}}{M_{red}} \Bigg)
\label{act_ov}
\end{equation}

\noindent where $A_a$ is the anode electrode surface area and $I_0$ is the exchange current density for mediator oxidation in reference conditions. Consequently, the output current can be estimated by: 

\begin{equation}
I_{cell}=\frac{(E_0-n_{conc}-n_{act})}{(R_{int}+R_{ext})} \frac{M_{red}}{\varepsilon+M_{red}}  
\label{icell}
\end{equation}

\noindent where $\varepsilon$ is a constant which bounds the output current at low values of $M_{red}$. While, Ohm's law provides the voltage produced:

\begin{equation}
V_{cell}=I_{cell} R_{tot}=I_{cell} \bigg( R_{int}+R_{ext} \bigg)
\label{ohms}
\end{equation}

\subsection{Biofilm formation}

The biofilm formation 
is an agent--based procedure simulating the attachment and growth 
of biomass. The attachment of biomass is simulated by a random selection in each time step of a predefined number of lattice cells ($k_{ata}$) to change their biomass concentration to a predefined initial value ($C^{bio}=C_0^{bio}$). The random selection of these cells is amongst the ones representing the front between the fluid domain and the anode electrode domain. 
Note that the initial value of the intracellular mediator in the cells selected for biomass attachment is set and equivalent to the biomass predefined initial value.

The growth of the biomass is following the double Monod equation providing the consumption rate (Eq. (\ref{monod})). Along with the growth of biomass, an analogous increase of the intracellular mediator is assumed. The spreading of the biomass is initiated when the biomass concentration in a lattice cell exceeds a predefined threshold ($C_{max}^{bio}$). The direction towards the spreading of the biomass happens, is randomly selected. If the under study cell is surrounded by cells representing unavailable space (solid boundaries and existing amounts of biomass) a random walk is executed to find a free cell towards where a displacement of biomass can occur. When a lattice cell representing an available choice is found, each cell with an existing value of biomass, transports its biomass value to the neighbour indicated by the random walk. Finally, when an adjacent cell of the initial cell exceeding the predefined threshold is freed up, a fraction of the biomass concentration of the initial cell ($fr_{spr}$) is subtracted from its biomass concentration and it is transported to the currently free cell. In addition to the transport of biomass being displaced by the random walk, the amounts of intracellular mediators follow the same procedure. 


Following the completion of the biofilm formation procedure, all the algorithmic steps of the model are repeated, starting with the calculation of fluid dynamics with LBM as formerly defined. The fluid flow is recalculated taking into account the updated biofilm as impermeable. Consequently, the advection and diffusion of the chemicals have to also be updated. Then, the electrical outputs are revised and, finally, the biofilm with the new concentrations and flow field is reformed. The model was implemented with Matlab R2016b.
%
%

\section{Results}
\label{sec4}

The applicability of the model can be verified with the comparison of the simulated results with measurements obtained in the laboratory. More specifically, here, the measurements obtained by an experiment with an innovative configuration of a single MFC cell, located onto a MFC system based on a building brick, were used. The MFC is comprised by the anode compartment accommodated in one cavity of the brick, the cathode compartment in another cavity and the ceramic semi-permeable separator between them.

The influnet 
of the anode compartment is considered for the simulation and depicted in the following figures to enter from the left side of the $x$-axis. Consequently, the effluent is considered to exit the anode compartment from the right side of the $x$-axis of the illustrated model configuration.

The values of the parameters used in the aforementioned equations that describe the functionality of the MFC are given in Table \ref{tab1}. These values are extracted by measurements in the laboratory experiment or adopted by previous attempts of describing MFCs and biofilms with mathematical equations in the literature \cite{bottero2013biofilm,pinto2010two}.

\begin{table*}[!tb]
\small
\begin{center}
 \begin{tabular}{| c | c | c |} 
 \hline
 Parameter & Description & Value \\ [0.5ex] 
 \hline\hline
 $X \times Y \times Z$ & Anode compartment dimensions & $17\times60\times65mm$ \\\hline
 $X' \times Y' \times Z'$ & Anode electrode dimensions& $17\times48\times65mm$ \\ 
  \hline
  $L_X\times L_Y$ & Model lattice dimensions & $60\times65$ \\\hline
 $\phi$ & Electrode porosity & 0.874 \\\hline
 $V$ & Fluid input velocity & $1.758\cdot10^{-2} mm/s$ \\ 
  \hline
 $v$ & Kinematic viscosity  &  $1.004 mm^2 / s$  \\\hline
  $\tau$ & Dimensionless relaxation time (LBM) &  0.6706 \\  
 \hline
 $D$ & Diffusion coefficient  &  $0.0012 mm^2 / s$  \\\hline
  $\tau_D$ & Dimensionless relaxation time (LBM) & 0.5036 \\  
 \hline
   $C^{in}_{s}$ & Input substrate concentration & 410 $mg \; substrate \; /\; L$ \\  
 \hline
 $Mtotal$ & Total amount of intracellular mediator  & 0.05 $\frac{mg\; mediator}{mg\; biomass}$ \\\hline
  $Y$ & Intracellular mediator yield & 0.5687 $\frac{mg\; mediator}{mg\; substrate}$  \\ 
  \hline
 $\gamma$& Mediator molar mass & 663400  $\frac{mg\; mediator}{mol\; mediator}$ \\\hline
 $m$& Electrons  provided  during reduction & 2 $e$ \\ 
  \hline
 $q_{max}$ & Maximum consumption rate & 8.48 $\frac{mg\; substrate}{mg\; biomass \;\cdot\; day}$ \\\hline
  $K_S$ & Half saturation constant for substrate & 20 $(mg \;substrate\;/\;L)$ \\ 
  \hline
 $K_{Mox}$ & Half saturation constant for oxidised mediator & $0.02\times Mtotal\; (mg\; mediator/L)$ \\\hline
  $I_{0}$ & Exchange current density & $0.001 \; A/m^2 $ \\
   \hline
    $E_{0}$ & Open circuit voltage & $0.7 \; V $ \\
   \hline
 $R_{ext}$ & External ohmic resistance (load) & 360  $\Omega$ \\\hline
 $R_{int}$ & Internal ohmic resistance & 360  $\Omega$  \\  
 \hline
  $k_{ata}$ & Predefined number of cells for attachment& 200   \\  
 \hline
 $C_0^{bio}$ & Initial biomass concentration & 450$mg\; biomass\; /\;L $  
 \\  
 \hline
 $C_{max}^{bio}$ & Threshold biomass concentration & 512.5 $mg\; biomass\; /\;L $  
 \\  
 \hline
  $fr_{spr}$ & Fraction of the biomass spreading & 40\%   \\  
 \hline
 
\end{tabular}
\end{center}
\caption{Parameters used in the execution of the model.}
\label{tab1}
\end{table*}

The procedures of calculating fluid dynamics and the advection--diffusion equation are running for enough time stems until an equilibrium is reached, 
while the bio-electrochemical model and the procedure of biofilm formation are updated once during each simulation interval. One iteration of the model corresponds to one hour of real time.

\begin{figure}[!h]
    \centering
       \includegraphics[width=0.5\linewidth]{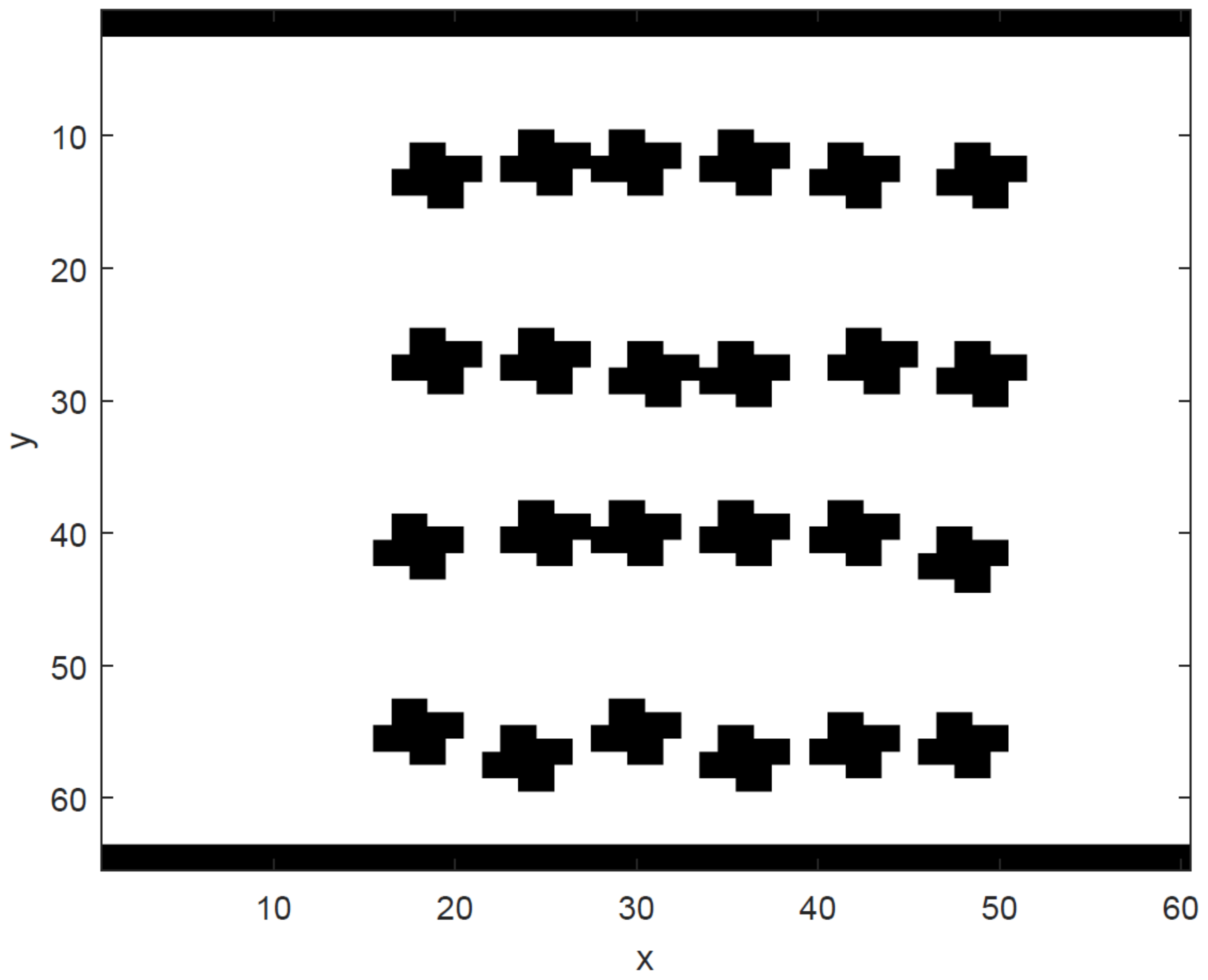}
    \caption{The geometry of the porous anode electrode.}
    \label{porous}
\end{figure}

In Fig. \ref{porous} the simulated anode compartment configuration is depicted. The black lines on top and on the bottom of the figure represents the impermeable boundaries (chassis) of the anode compartment. The influent is entering the compartment from $x=0$ and the effluent is exiting from $x=60$. Whereas, the porous medium of the anode electrode is illustrated with the black shapes that are repeated throughout the figure. Note that a random geometry of the anode electrode was adopted. Nonetheless, this selection imposes no limitation on the functionality of the model. The formation of the geometry of the anode electrode can be disorganised, lacking a regular geometry or order, or based on mathematical self-organising mathematical tools \cite{bandman2011using}.

\begin{figure}[!h]
    \centering
    \includegraphics[width=0.5\linewidth]{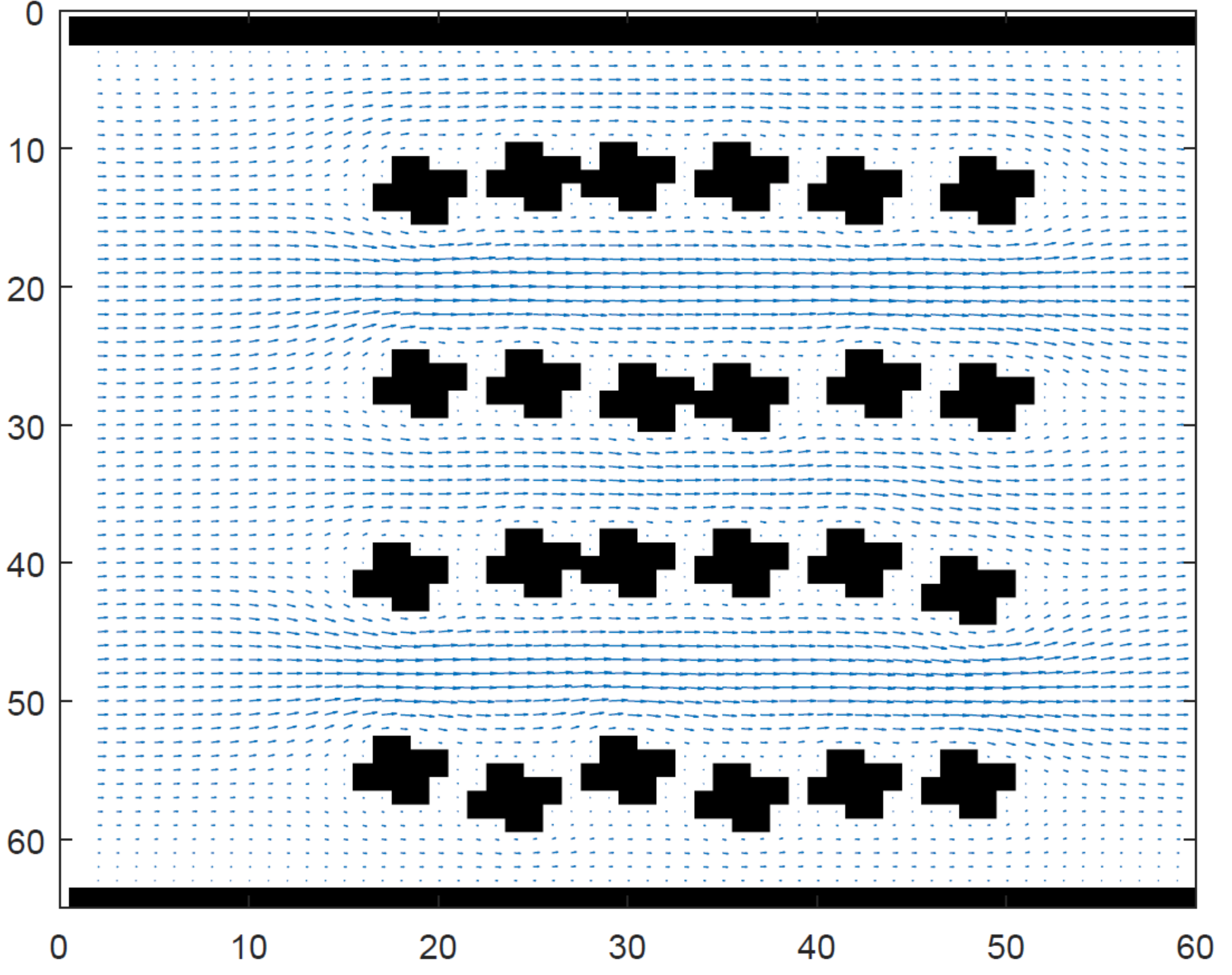}
    \caption{Flow field in the anode without a developed biofilm.}
    \label{flow1}
\end{figure}

\begin{figure}[!h]
    \centering
    \includegraphics[width=0.5\linewidth]{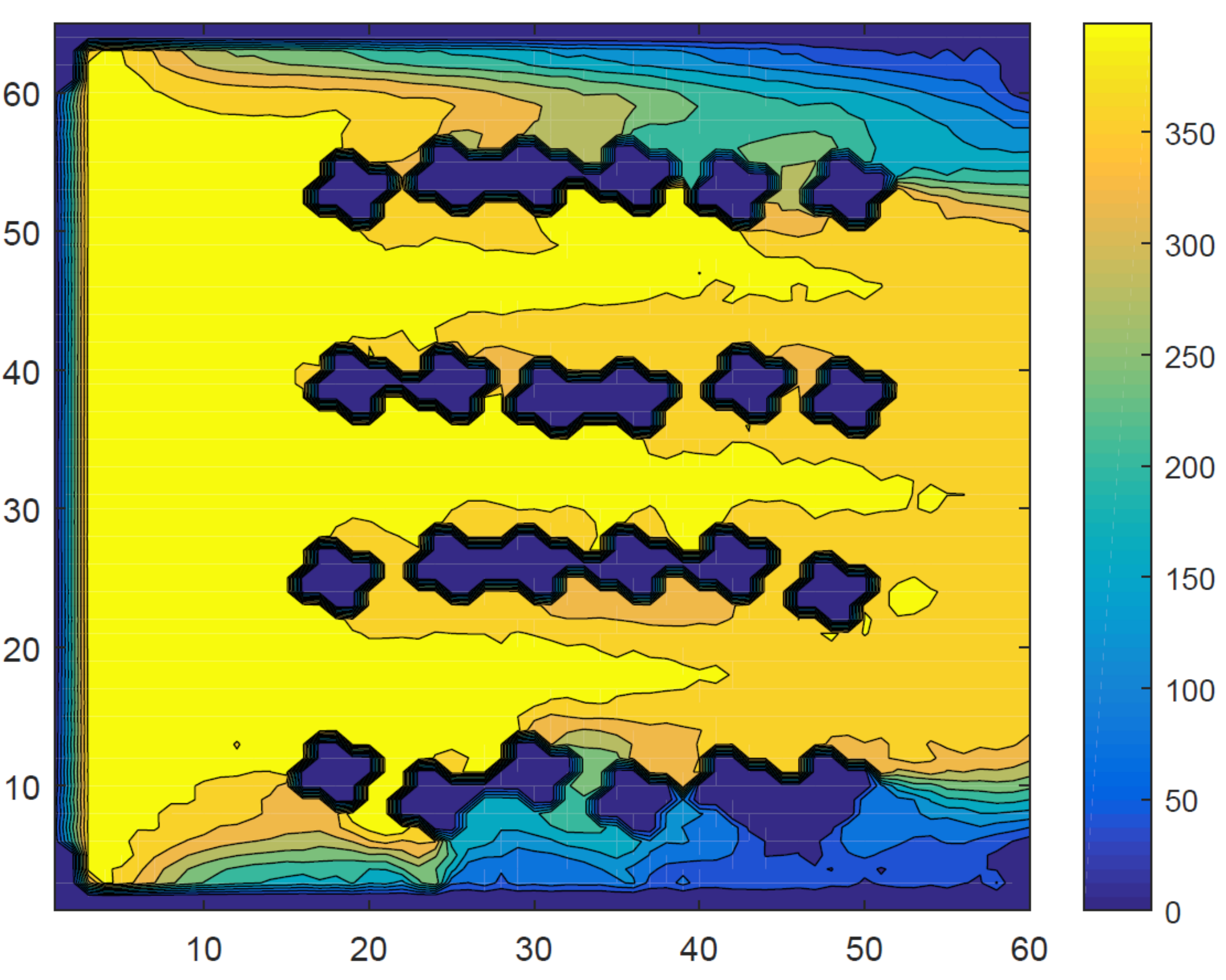}
    \caption{Concentration field in the anode without a developed biofilm (in $mg/L$).}
    \label{conc1}
\end{figure}

The flow and concentration field in the anode compartment can be calculated without any biofilm formed on the anode electrode. The results of such calculations are illustrated in Figs. \ref{flow1} and \ref{conc1}.

\begin{figure}[!h]
    \centering
       \includegraphics[width=0.5\linewidth]{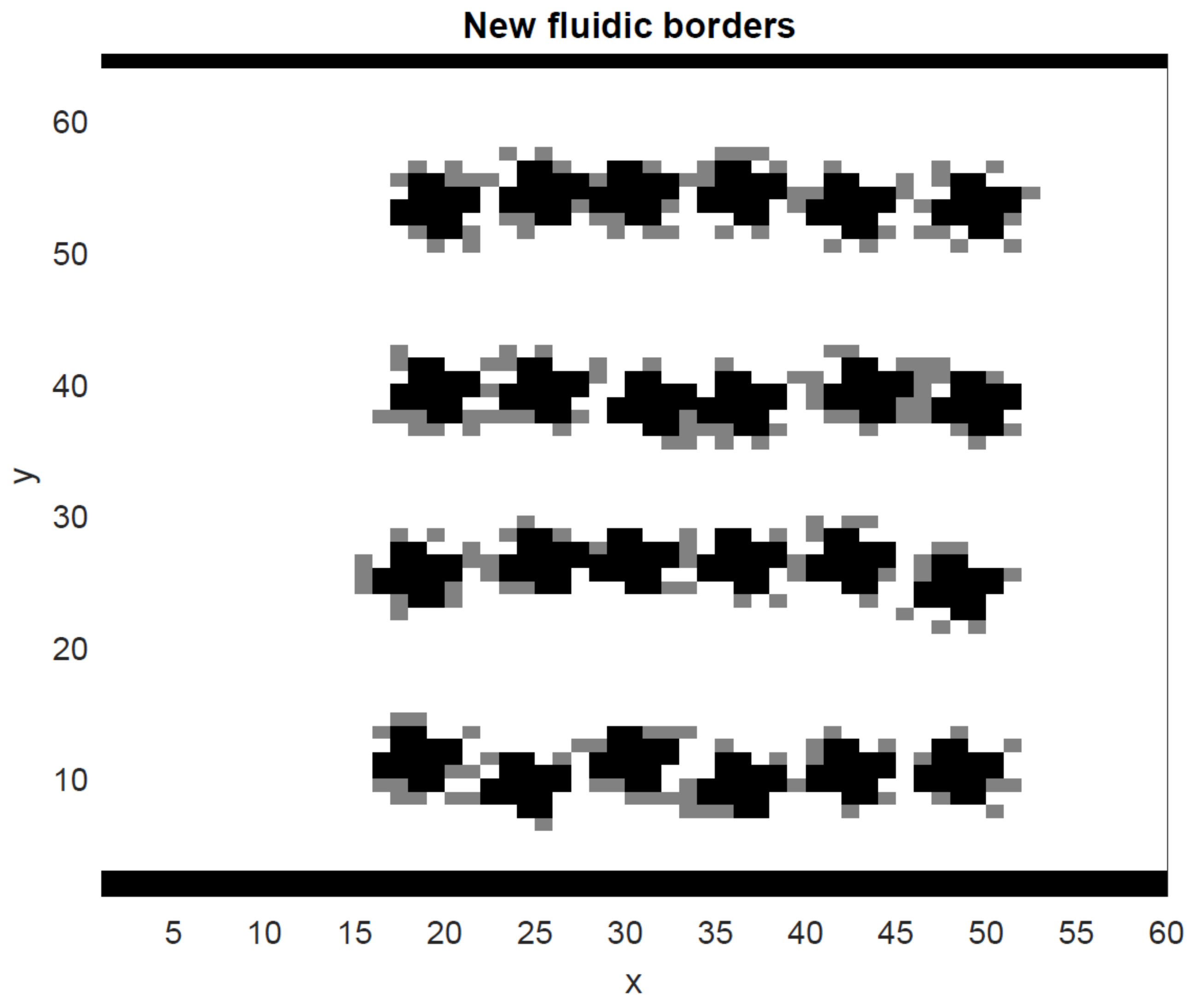}
    \caption{The configuration of the initial biofilm on the anode electrode porous medium.}
    \label{biofilm}
\end{figure}

\begin{figure}[!h]
    \centering
       \includegraphics[width=0.5\linewidth]{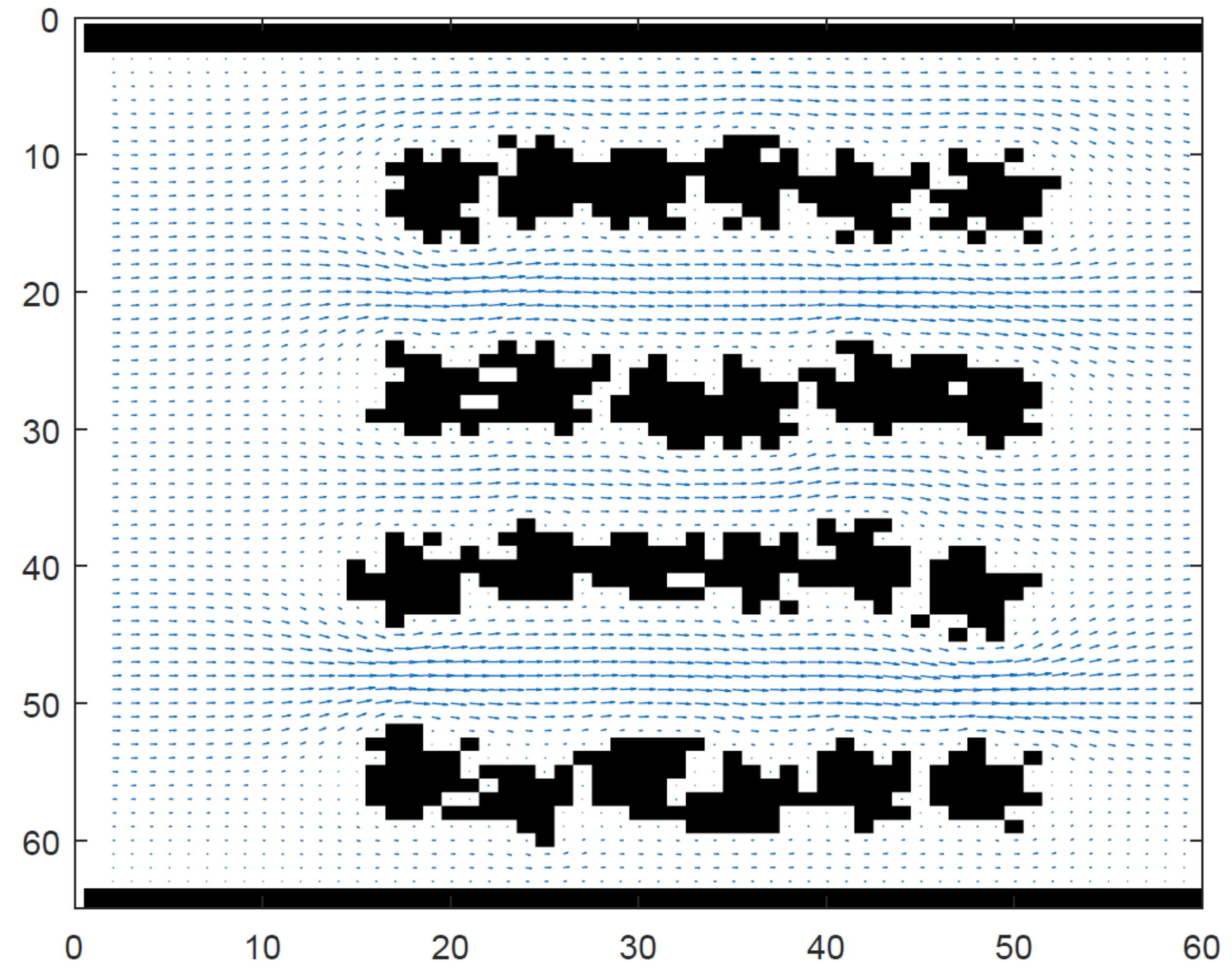}
    \caption{Flow field in the anode after the development of the initial biofilm.}
    \label{flow2}
\end{figure}

\begin{figure}[!h]
    \centering
       \includegraphics[width=0.5\linewidth]{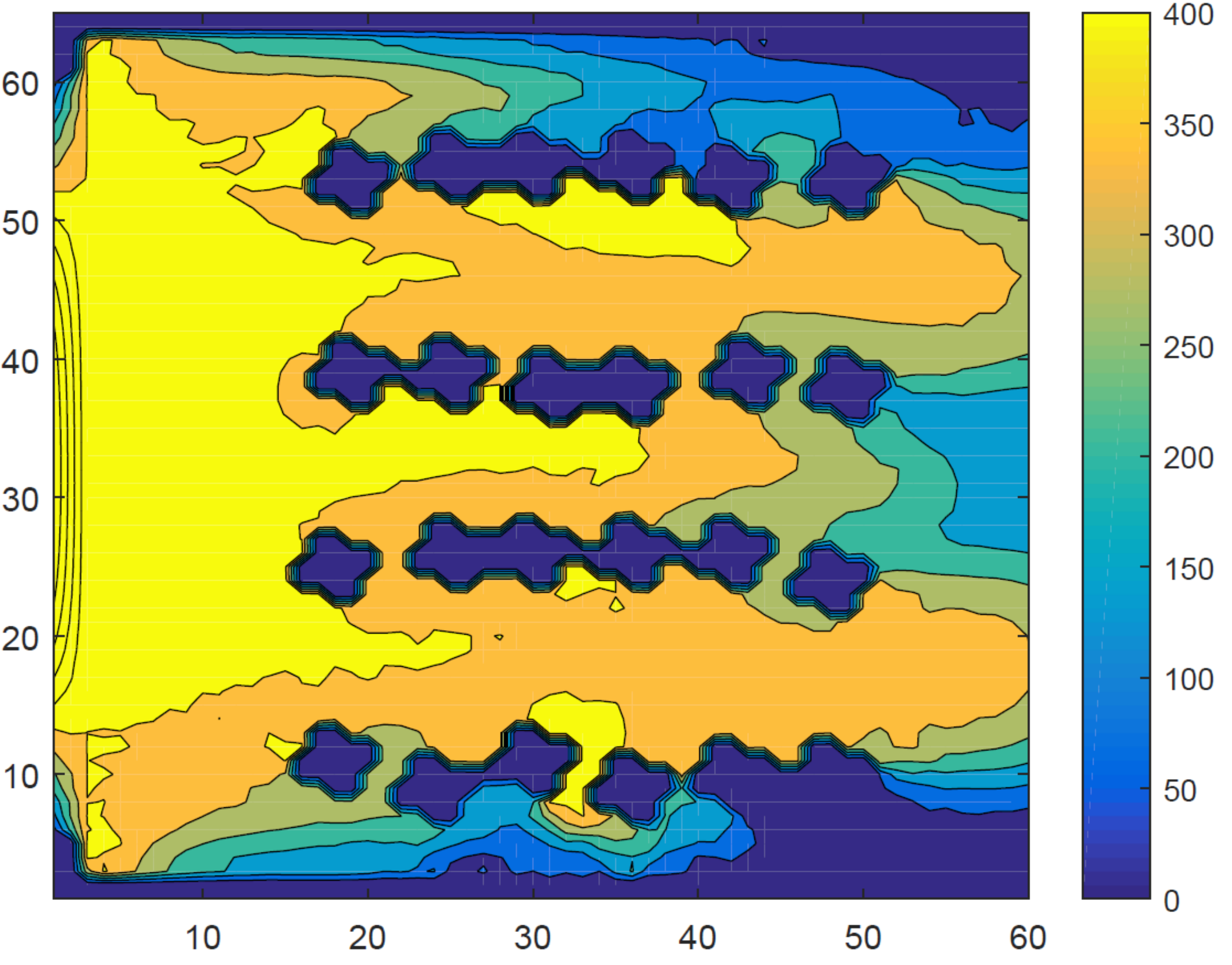}
    \caption{Concentration field in the anode after the development of the initial biofilm (in $mg/L$).}
    \label{diff2}
\end{figure}

When a biofilm is developed on the anode electrode (illustrated in Fig. \ref{biofilm} as the gray spots attached on the black shapes that represent the porous electrode) then the flow and concentration fields are recalculated taking into consideration the new configuration. The results of the flow field LBM calculation are depicted in Fig. \ref{flow2} and the results of the advection-diffusion LBM are depicted in Fig. \ref{diff2}.

\begin{figure}[!h]
    \centering
       \includegraphics[width=0.5\linewidth]{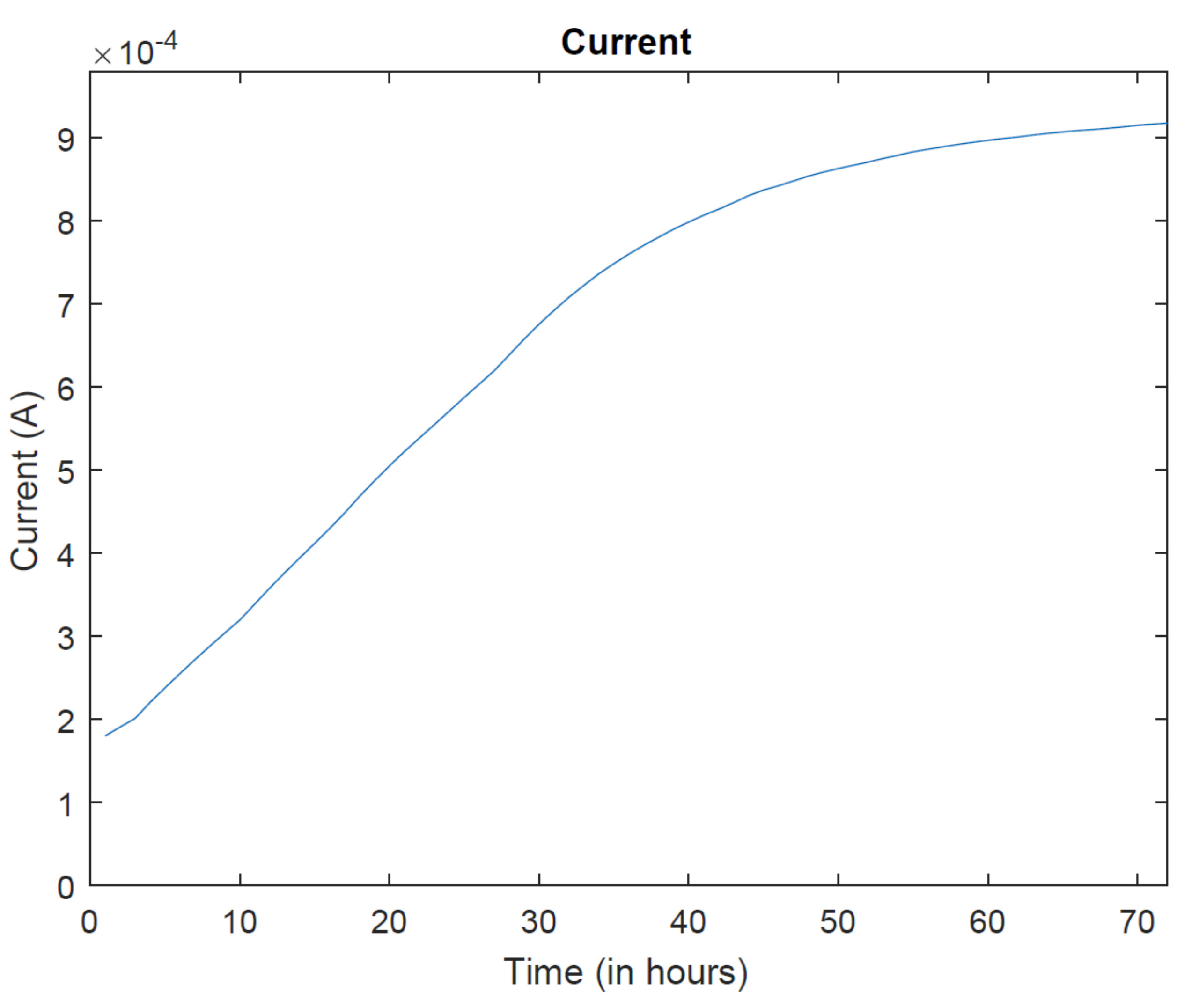}
    \caption{MFC output current over time in hourly intervals.} 
    \label{current}
\end{figure}

Finally, the simulation of the functionality of the under study MFC is carried out for a simulation of 72 hours and the results are presented in the following figures. In Fig. \ref{voltage} the output voltage of the MFC is illustrated. The output voltage is asymptotically increasing during the first 60 hours and then reaches a steady state of approximately 330 mV that is in agreement with the laboratory based measurements. Moreover, the output current depicted in Fig. \ref{current} is following the increase of the voltage curve and it reaches a steady value of 0.91 mA (similar with the laboratory experiment).

\begin{figure}[!h]
    \centering
       \includegraphics[width=0.5\linewidth]{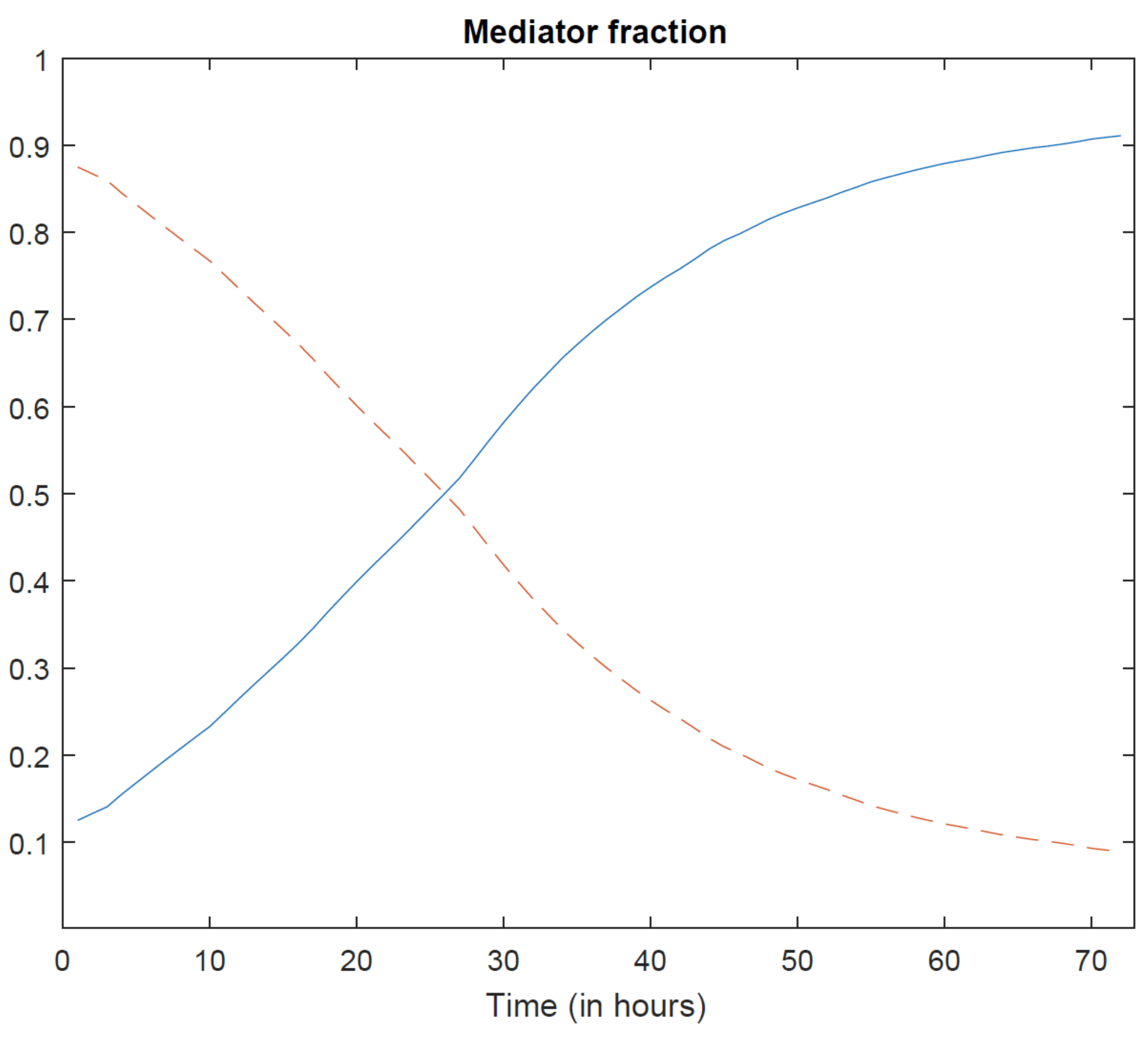}
    \caption{Intracellular mediator concentrations fractions over time in hourly intervals (reduced -- solid line, oxidised -- dashed line).}
    \label{mediators}
\end{figure}

In Fig. \ref{mediators} the ratio of the reduced and the oxidised mediator to the total amount of intracellular mediator are depicted through time. The concentration of reduced mediator (solid line) is following the evolution of the output voltage, whereas the concentration of the oxidised mediator (dashed line) is following a complete opposite asymptotic curve of the reduced mediator, as derived by Eq. (\ref{med1}).

\begin{figure}[!h]
    \centering
       \includegraphics[width=0.5\linewidth]{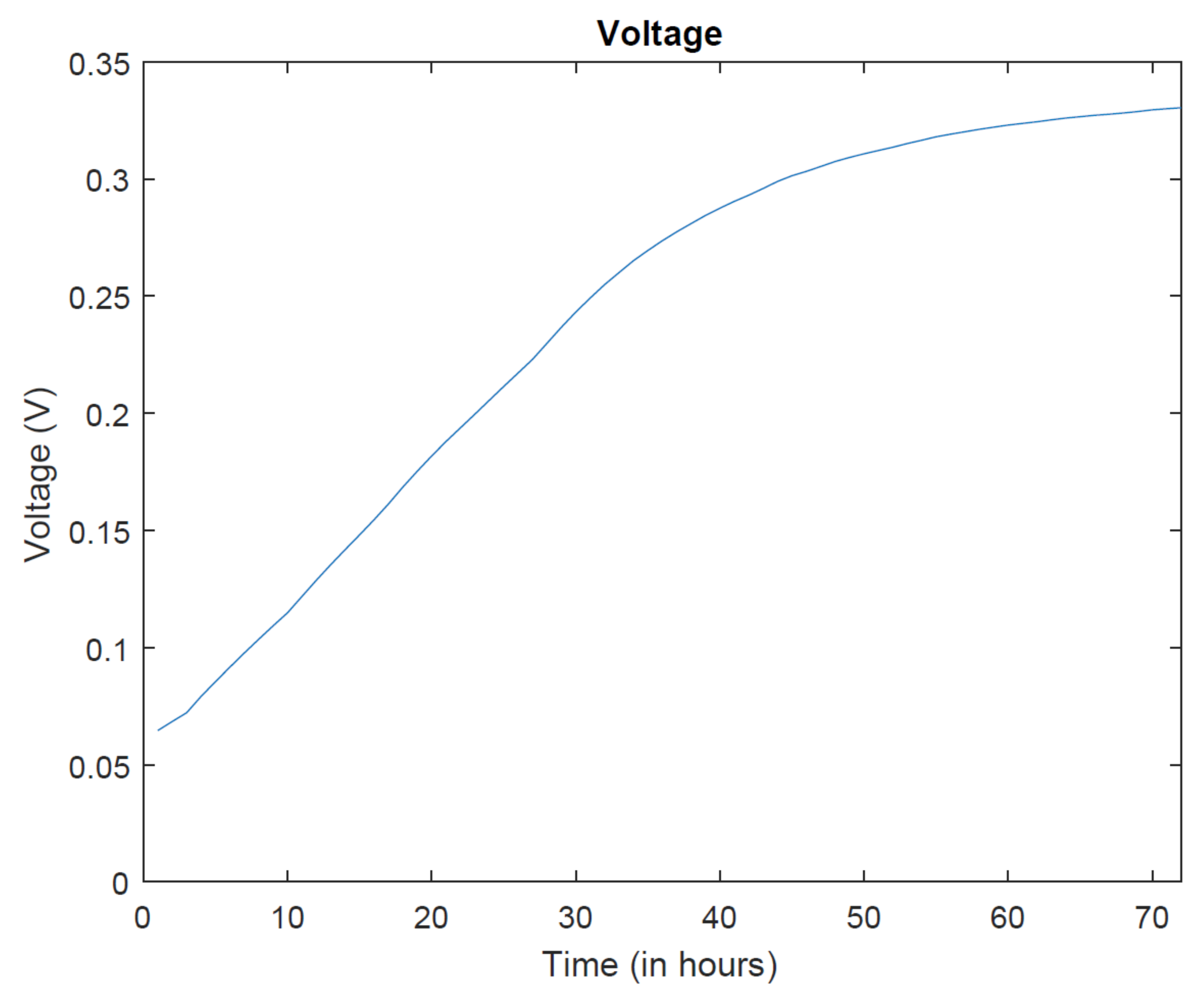}
    \caption{MFC output voltage over time in hourly intervals.} 
    \label{voltage}
\end{figure}

%
%

\section{Conclusions}
\label{sec5}

We proposed a model of MFC that incorporate geometry of the electrodes. The model implements two LBMs to calculate the fluid flow field and the advection--diffusion phenomena of chemicals inside the anode compartment. Also, an agent--based model calculates the attachment and spreading of biomass on the electrode. The output voltage is calculated using bio-electrochemical processes.

An advantage of the model is that it is designed to consider the flow dynamics of the influent and the advection--diffusion phenomena of chemicals in comparably short time intervals (seconds and minutes). Thus, the model can be utilised in predicting transitions of the behaviour of a MFC from one state to another in this time span when the relevant parameter, namely influent velocity or fuel concentration is altered.

We adopted a random geometry of the anode electrode. This imposes no limitations on the model's functionality. The formation of the geometry of the anode electrode can be disorganised, lacking a regular geometry or order, or based on mathematical self-organising mathematical tools \cite{bandman2011using}. Studying particular geometries will be a topic of future research. This could include analysis of the effect of specified geometries and micro-structures on the development of biofilms in a MFC and its electrical and chemical outputs, e.g. geometries based data from X-ray tomographic microscopy \cite{prasianakis2013simulation}.

The biofilm formed during modelled development of a MFC is assumed to impose a very high resistance to fluid dynamics, thus, the speed of fluid components throughout the biofilm was considered negligible. However, this is not an ideal representation of a real biofilm. More work would be required to take into account the permeability of biofilms and communication between cells in the matrix.

While, in the simulated experiments the total concentration of the intracellular mediator is regarded as constant and analogous to the biomass concentration, in reality, fractions of oxidised and reduced forms of the mediator can vary. This could be another aspect of future work.

  \section*{Acknowledgment}

This work was supported by the European Union's Horizon 2020 Research and Innovation Programme under Grant Agreement No. 686585.

\end{document}